# Adhesive contact of a compliant sphere to an elastic coated substrate: the thin film limit

E. Barthel

Surface du Verre et Interfaces, CNRS/Saint-Gobain, BP 135, F 93303 Aubervilliers Cedex France

## *Abstract*

Experimental results for adhesive contacts on substrates coated with elastomeric thin films have recently been obtained by C. Tardivat and L. Léger (J. Adhesion Sci. Technol. **15** 1055-1078 (2001)) by the so-called JKR test, which provides both adhesion energy and elastic modulus. These data show that on substrates coated with thin films the adhesion and effective elastic modulus of the sphere depend upon the film thickness. In keeping with the experimental conditions, we try to interpret these data using a simple model (E. Barthel and A. Perriot, J. Phys. D 40, 1059–1067 (2007)) in the thin film limit, *ie* when the film thickness is small compared to the contact radius. Although the film does impact the *local* crack tip stress field, we show that no effect on the *macroscopic* contact variables is expected for the adhesion to coated substrates in such confined geometries. The deviations from the experimental results are ascribed to the idealized contact boundary conditions assumed in the model.

## *Introduction*

A good understanding of the adhesive contact to coated substrates is required for a wide range of applications in areas like surface modification, paints and adhesives. However, few reliable data on adhesive contacts in the presence of thin elastic films can be found in the literature. One notable contribution in the area is an early paper by C. Tardivat and L. Léger (TL) [1]. Using a JKR test with a soft polyisoprene (PI) lens contacting a thin PI layer on silicon, they gathered comprehensive information, mainly under the form of force *vs.* contact radius curves. They probed the impact of film thickness and film modulus on the effective adhesion and contact stiffness as measured in a JKR test. Well aware of the difficulties incurred in modelling adhesive contacts to coated substrates, TL took care to use adequately reduced variables to assess the deviation of their data from the expected JKR (ie adhesive contact on a homogeneous half-space) model [2] and provide for a more straightforward discussion.

The aim of the present contribution is to compare these experimental results with the simplest exact extension of the JKR model to coated substrates. In this extension, the contact between film and indenter is completely frictionless, the adhesion is purely local and the polymer perfectly linear elastic. These are clearly approximations from which we gain the possibility to apply a simple linear elastic model for the adhesive contact on coated substrates [3, 4]. We show that, given the high degree of confinement experienced by the coating in the TL experiments, no deviation from JKR should appear due to the film although the crack tip stress field is affected by the additional thin film compliance. The experimental results, which show the contrary, suggest that more complex boundary conditions such as contact with friction are needed for accurate modelling of the TL data.

## *Adhesive contact to a homogeneous half-space: JKR on a bulk*

The standard JKR theory [2] is used by TL to analyze their data as force *vs.* contact radius plots in the following reduced form:

$$\frac{F}{\sqrt{6\pi}a^{3/2}} = \frac{4E^*}{3\sqrt{6}}\left(\frac{a^{3/2}}{\sqrt{\pi R}}\right) - \sqrt{\frac{4wE^*}{3}} \qquad (1)$$

where $F$ is the force, $a$ the contact radius, $R$ the sphere radius, and $w$ the adhesion energy (Fig. 1a). In the JKR model, the compound reduced elastic modulus $E^*$ is defined by $E^{*-1} = E_2^{*-1} + E_0^{*-1}$ with $E_i^* = E_i/(1-\nu_i^2)$. Experimentally, plotting $F/\sqrt{6\pi}a^{3/2}$ as a function of $a^{3/2}$ should result in a straight line if the contacting half-spaces are indeed homogeneous. Such a straight line was obtained by TL for a contact on bare silicon with tuned chemistry but no film [1]. In such instances, this reduced plot – let us call it a JKR plot - is a good way to obtain the two experimental parameters: reduced elastic modulus and adhesion from slope and intercept as suggested by Eq. 1.

## *Adhesive contact to a coated substrate: JKR on a film*

### Experimental results

It is expected that deviations will appear for non homogeneous half spaces. Such deviations were indeed observed for contacts to coated substrates [1]. However, for contact radii significantly larger then the film thickness, it appears that the curves were very close to linear, so that a JKR plot would still yield a slope and intercept and therefore an effective modulus and effective adhesion energy. Only at rather small contact radii did the reduced force-radius plot deviate from the expected straight line.

Now what should be noticed here is that the materials are soft (a few MPa), the adhesion non negligible (tens of mJ/m$^2$) and the sphere radius large (1 mm) so that the typical contact radius (50 microns) under purely adhesive loading and no external force far exceeds the layer thickness (1-10 microns) considered in the experimental study: we specifically deal with the question of adhesion on highly confined films. We expect from thin film theory that under such circumstances, the confined film does not affect the contact mechanics which should be ruled by the bulk materials (sphere and substrate) [5]. However, the TL experimental results point to the contrary.

More specifically, we will here consider the data where a soft sphere ($4E_2^*/3 = 1.95$ MPa) was pressed against a barely less compliant film ($E_1^* = 3.5$ MPa) deposited on a rigid substrate (silicon) (see Fig. 1b for notation). Cursory investigation of the other case described in TL does not contradict the conclusions listed below. The TL data analysis shows that the measured equivalent modulus is slightly reduced (10%) compared to the sphere modulus measured directly on a rigid substrate, independently from the film thickness, while the adhesion energy is constant for films thicker than 3 microns but, strikingly, gradually decreases by up to about 25% for film thicknesses below about 3 microns. Of course while the no-film limit case yields the reference (sphere) reduced modulus, the adhesion energy measured in this limit case cannot be used to analyze the rest of the data because the surface chemistry is different.

**Modelling the contact**

Three different approaches have been developed to model the adhesive contact to coated substrates. Based on finite element (FE) calculations, Sridhar and Johnson on the one hand [6], and Shull on the other hand [7] have calculated adhesive contact behaviours on coated substrates. In particular, Sridhar has shown that his data could model the TL results [8].

Along a different line, the model used in the present paper was developed as the natural extension of our previous approach to the (adhesionless) contact to coated substrates [5, 9].

However, the previous developments [3, 4] considered only a rigid contacting body and this must now be replaced by a soft, elastomeric one. One of the additional virtues of this model is that such extension is direct, as explained below.

## *Model*

### Contact to a coated substrate

Along the lines of the JKR methods, the core of our approach [4] is to express the solution as the linear superposition of an adhesionless contact and a flat punch. As an exemple, the force is

$$F = F_0 + S\delta_{fp} \qquad (2)$$

where $F$ is the total force, $F_0$ is the force for the adhesionless contact, $S$ the contact (ie flat punch) stiffness and $\delta_{fp}$ the flat punch penetration.

To apply this strategy, the normal surface stress distribution under the contacting punch is calculated through the use of adequate transforms of the normal surface stress $g(r)$ and normal surface displacement $\theta(r)$. For homogeneous linear elastic half spaces, equilibrium results in a proportionality relation between $g(r)$ and $\theta(r)$, to wit

$$\theta(r) = \frac{2}{E^*} g(s)$$

For thin films, the more elaborate relation

$$\theta(r) = \int_0^a C(r,s)g(s)ds \qquad (3)$$

must be used where for a rigid punch $C(r,s)$ depends upon the geometry and mechanical parameters of the coated substrate [5]. The benefit of the $g(r)$ and $\theta(r)$ transforms is to keep the mathematical complexity to a minimum.

For the adhesionless sphere, after normalization, the force $\Pi_0$, defined from $F_0$ by

$$F_0 = \frac{a^3 E_2^*}{2R} \Pi_0 \quad (4)$$

can be calculated (Fig. 2). Further, under flat punch boundary conditions, the contact radius is fixed and independent from the penetration. However a stress singularity of amplitude $g(a) \neq 0$ appears at the contact edge. Looking at the adhesive contact as a crack problem, $g(a)$ is found to be proportional to the stress intensity factor at the crack tip. In the full adhesive contact it reflects the adhesion and is the origin of the cusp typically observed at the contact edge of the adhesive contact of soft bodies (Fig. 1a).

Calculating the stress distribution for a given penetration also provides the normalised contact stiffness $E_{eq}$ defined from the contact stiffness $S$ by

$$S = 2a E_2^* E_{eq} \quad (5)$$

and the relation between the amplitude of the singularity $g(a)$ and penetration $\delta_{fp}$

$$g(a) = \frac{\delta_{fp} E_2^*}{2} \Gamma \quad (6)$$

where $\Gamma$ is the normalized proportionality factor (Fig. 3).

For our purpose, a modification to our previous algorithm [4] is required by the fact that the punch is no longer rigid. Fortunately, due to the frictionless boundary conditions, this can be included in an *exact* and simple way by adding a diagonal $1/E_2^* \delta(r-s)$ term to the response function $C(r,s)$ in Eq. 3. This contrasts to the Finite Element calculations where a full punch description would be required and must be circumvented by an approximate scheme [8].

With this in mind, and to easily recover the soft sphere limit case, we also modified the normalization compared to our previous paper: all the quantities homogeneous to stress are now normalized to the sphere reduced modulus $E_2^*$ (as in Eqs. 4-6) instead of the film reduced modulus $E_1^*$, which means that values computed with our previous algorithm must be multiplied by $E_1^*/E_2^*$.

As a result of the extension to a compliant sphere, however, the number of parameters in the model increases significantly, and the exploration of the full parameter space even in reduced form is impractical. Therefore, we will present typical results for experimental parameters similar to those used in the TL investigations. In particular, in agreement with the elastomeric nature of the film, we model the case of an incompressible film ($\nu_1 = 0.5$). Incompressibility has previously been shown to affect the overall elastic response dramatically [9, 10], with considerable impact on the rigidity of the confined film.

Calculations were performed with the parameters summarized in Table 1 for $a/t$ ratios ranging from 0.001 to approximately 1000, where $t$ is the film thickness. The resulting evolution of adhesionless contact force $\Pi_0$, equivalent modulus $E_{eq}$ and stress intensity factor $\Gamma$ as a function of $a/t$ are plotted on Figs. 2 and 3.

**Limit cases**

As expected, at small $a/t$, the mechanical response is dominated by the compound sphere-layer modulus

$$E_s^{*-1} = E_1^{*-1} + E_2^{*-1} \qquad (7)$$

The limit values are $\frac{8}{3} E_s^*/E_2^*$ for the normalized force and and $E_s^*/E_2^*$ for the normalized reduced modulus and normalized stress intensity factor. The limit values at large $a/t$, where the mechanical response is dominated by the sphere compliance are 8/3 for the normalized force, 1 for the normalized reduced modulus, and $\sqrt{E_s^*/E_2^*}$ for the normalized stress intensity factor. This limit for $\Gamma$ has previously been evidenced on the ground of energy considerations [4, 7] such as also developed below (Eq. 8). This latter value demonstrates how the local field at the contact edge is affected by the increased compliance due to the film while the macroscopic (integral) response such as the force and stiffness remain unchanged from the homogeneous substrate case. This is similar to the thin film limit in a double-cantilever beam configuration [11].

**Adhesive contact**

From these reduced quantities, and using the contact parameters from Table 2 and the actual sphere reduced modulus $E_2^*$, one can calculate the adhesive contact solution.

The only additional detail needed is the local crack tip relation between adhesion energy and stress intensity factor. This relation is not an intrinsic part of the adhesive contact calculation and must be specified: different additional dissipation mechanisms could be injected here if needed (small scale yielding, crack tip viscoelasticity). Within the elastic framework, the usual assumption is that the crack tip is dominated by the *local* elastic response so that [3, 4, 6-8] the relevant compliance at the crack tip is the compound sphere-layer modulus (Eq. (7)) and

$$\frac{2g(a)^2}{\pi a} E_s^{*-1} = w \qquad (8)$$

Then in normalized form, Eq. 1 expands into

$$\frac{F}{\sqrt{6\pi}a^{3/2}} = \frac{E_2^*}{2\sqrt{6}}\left(\frac{a^{3/2}}{\sqrt{\pi R}}\right)\Pi_0 - \sqrt{\frac{4wE_s^*}{3}\frac{E_{eq}}{\Gamma}} \qquad (9)$$

Force curves for different film thicknesses are displayed on Fig. 4. They are directly calculated from the previous results (Fig. 2 and 3) with the parameters from Table 2. As in [4], we insist that the subsequent calculations, for various film thicknesses and adhesion energies are derived directly from the previous numerical results ($\Pi_0$, $E_{eq}$, $\Gamma$) by the elementary arithmetics of Eq. (9) without any additional involved numerics. In addition, two JKR limits have been plotted. One is the JKR curve for an infinitely thick film (sphere on film). This results in a smaller effective adhesion and a smaller slope than the no-film (sphere on substrate) JKR plot also shown on Fig. 4.

For all finite film thicknesses, the force curves converge to the JKR (sphere-film) curve at small $a/t$, as expected from the insertion of the limiting values stated above (section limit cases) into Eq. 9. However, for all practical purposes, this part of the curve is insignificant. The practical result which appears in Fig. 4 is that for film thicknesses significantly smaller than the zero force contact radius (let us say here less than 10 µm), no significant deviation (less than 3%) from the JKR (sphere) model is expected. This applies both to the stiffness calculated from the slope of the plot and the adhesion energy calculated with the sphere stiffness. These results directly show that although the local stress field is affected and now depends upon the compound stiffness, no direct impact on macroscopic variables is expected in this highly confined film limit.

## *Discussion*

The non-linear relation between the JKR plot parameters on the one hand and elastic modulus and adhesion energy on the other hand makes it difficult to accurately interpret the deviations between our predictions and the TL data. From the present analysis, the 10% reduction of the effective modulus is not expected, nor the decrease of the adhesion for very thin films. In this respect it would be interesting to investigate in more detail the agreement with the experimental data found by Sridhar et al. [8]. A JKR plot could be useful for that purpose.

Several phenomena not included in the present model may play a role. A possibility is the impact of the friction or sliding conditions at the interface, which is neglected here. One could also suspect that the effective modulus felt by the interaction zone is larger for thin films if the size of the interaction zone is not actually very small. This however would presumably result in an overestimation of the adhesion energy because an erroneously small value of the modulus would be used to extract *w* from Eq. 9. This is not consistent with the deviations observed by TL.

However, this idea points in the direction of the impact of the elastic film on the local crack tip response. Although a thin confined film will not affect the relations between macroscopic contact variables and the bare JKR result is obtained, it does affect the stress intensity factor as shown by Eqs. 6 or 8. This means that the film will impact on the crack tip mechanics at least by the additional compliance. Any small scale yielding or viscoelastic crack tip process will be tuned by the elastic response of the film. In such a case, this local field effect will propagate to the macroscopic scale as an effective adhesion due to additional dissipative processes.

## *Conclusion*

We have shown how the elastic contact to coated substrates algorithm can be extended to compliant spheres. The thin film limit has been investigated and we have shown that as expected, at the macroscopic scale, the JKR model holds when the film thickness is small compared to the contact radius. However, the local stress field at the crack tip is affected.

As a result, in the TL experiments, the JKR plot is expected to provide the sphere modulus and the adhesion energy. The deviations observed in the experiments are probably due to friction effects at the interface which are not taken into account in the present model.

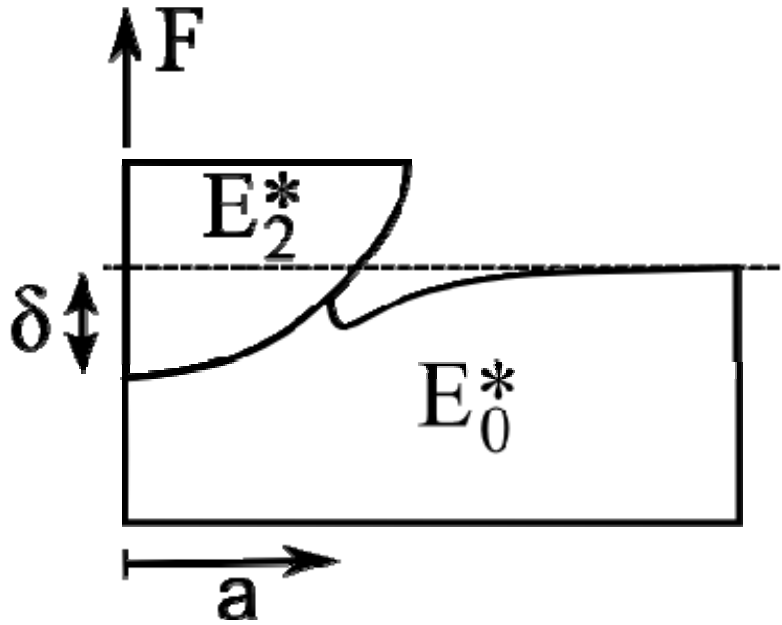

Fig. 1a: Adhesion of an elastic sphere to a homogeneous substrate.

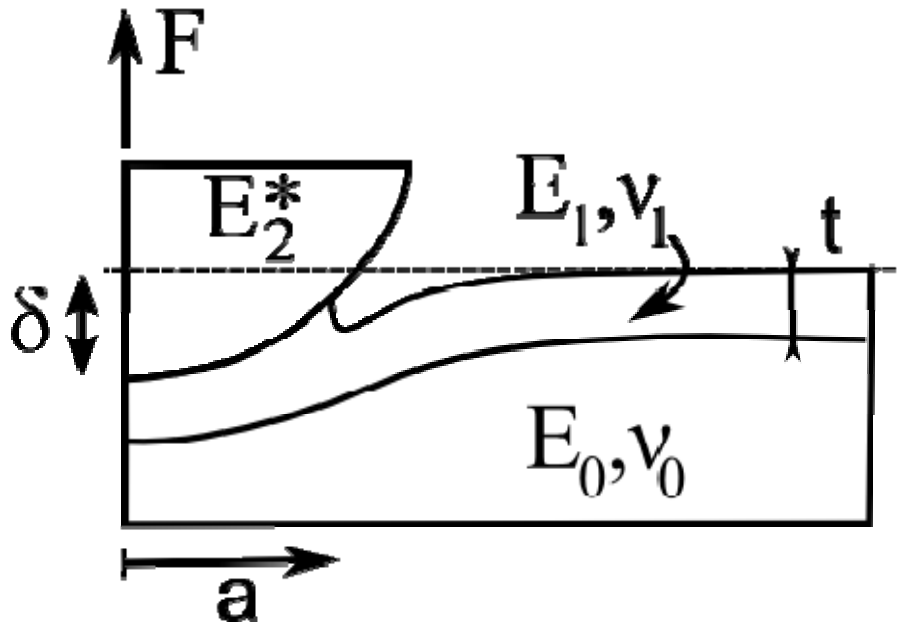

Fig. 1b: Adhesion of a sphere to a coated substrate. The adhesive contact is primarily ruled by the relation between penetration $\delta$ (macroscopic contact variable) and stress intensity factor (local crack tip field, proportional to $g(a)$, where $a$ is the contact radius).

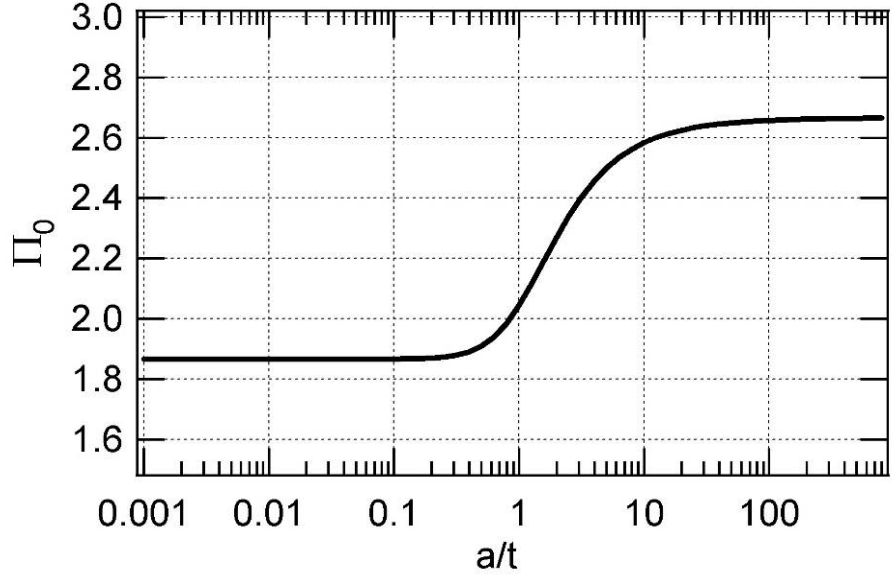

Fig. 2: Normalized contact force $\Pi_0$ for an adhesionless sphere on a coated substrate for the values of Table 1.

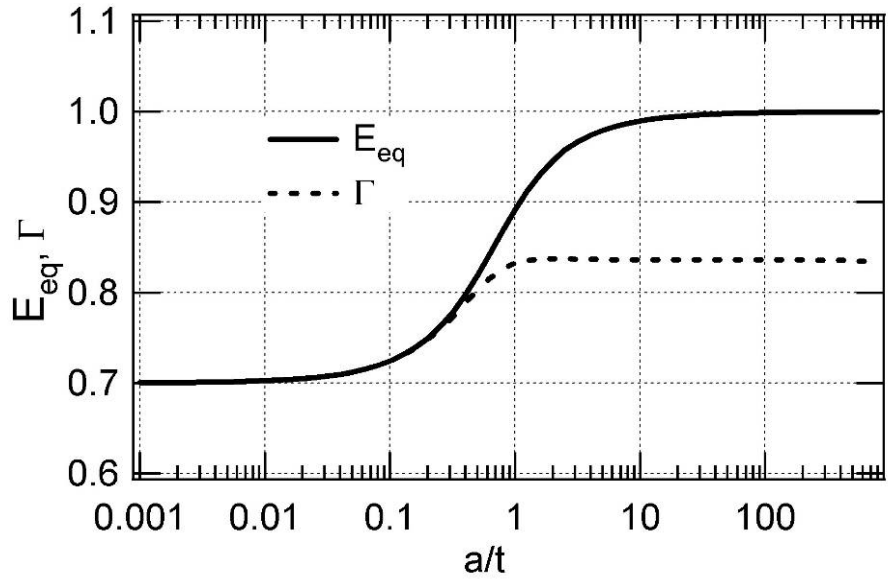

Fig. 3: Normalized equivalent modulus $E_{eq}$ (normalized contact stiffness) and stress intensity factor $\Gamma$ (normalized stress field singularity) for an adhesionless sphere on a coated substrate with the parameters from Table 1. For these values, $E_s^*/E_2^* = 1/(1+\tfrac{3}{7}) = 0.7$.

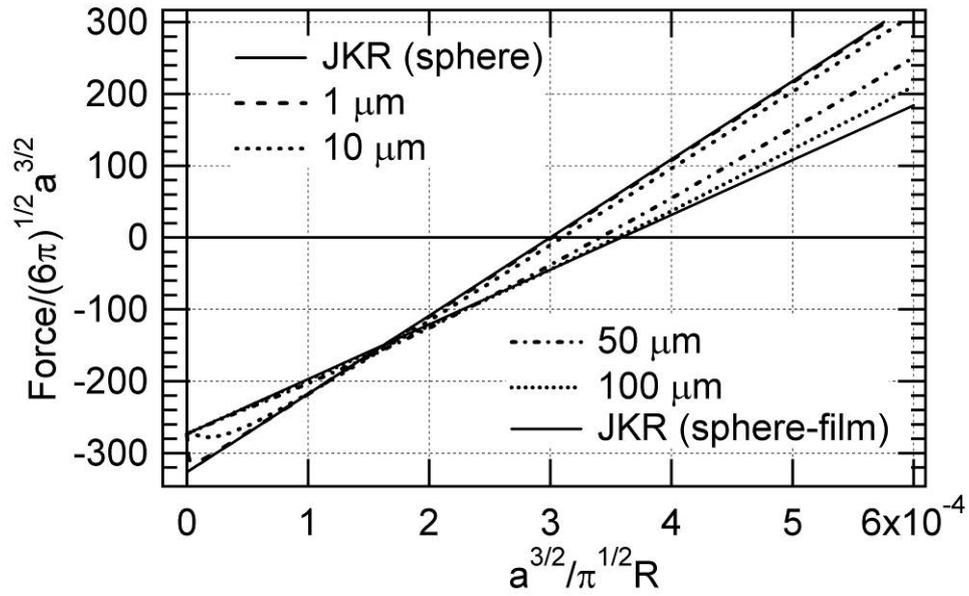

Fig. 4: Reduced JKR plots for various film thicknesses, with the parameters of Tables 1 and 2. Also shown are the JKR asymptotes for a rigid substrate without coating (sphere) and an infinitely thick coating (sphere-film). The JKR (sphere) curve is nearly always coincident to the 1 μm thick film curve, except at very small contact radii. The JKR (sphere-film) curve is coincident to the 100 μm thick film curve except at larger contact radii.

| Sphere reduced modulus (MPa) | Film modulus (MPa) | Film Poisson ratio | Substrate modulus (MPa) | Substrate Poisson ratio |
|---|---|---|---|---|
| $E_2^*$ | $E_1$ | $\nu_1$ | $E_0$ | $\nu_0$ |
| 2.0 | 3.5 | 0.5 | $2.10^5$ | 0.25 |

Table 1: Material parameters for calculated normalized variables displayed on Figs. 2 and 3. The moduli were normalized to $E_2^*$ for these calculations.

| Adhesion energy (J/m$^2$) | Sphere radius (mm) |
|---|---|
| $w$ | $R$ |
| 0.040 | 1.0 |

Table 2: Material parameters for the calculation of adhesive contact results displayed on Fig. 4.